\documentstyle[manuscript,aps]{revtex}
\begin{document}
\draft
\title{Some Remarks on the Bel--Robinson Tensor\footnote{This paper is
an amended version of an Essay which received a "honorable mention" in
the 1999 Essay Competition of the Gravity Research Foundation.}}
\author{Janusz Garecki\footnote{E-mail: garecki@wmf.univ.szczecin.pl}\\
Institute of Physics, University of Szczecin\\
Wielkopolska 15, 70--451 Szczecin, POLAND}
\date{\today}
\maketitle
\begin{abstract}
In this paper we present our point of view on correct physical interpretation of the
Bel--Robinson tensor within the framework of the standard General
Relativity ({\bf GR}), i.e., within the framework of the {\bf GR} without
supplementary elements like arbitrary vector field, distinguished tetrads field
or second metric. We show that this tensor arises as a consequence of the
Bianchi identities and, in a natural manner, it is linked to the differences of
the canonical gravitational energy--momentum calculated in normal coordinates
{\bf NC(P)}.
\end{abstract}
\newpage
\section{Introduction}
In analogy to the symmetric energy--momentum tensor of the electromagnetic
field, the Bel--Robinson superenergy tensor is defined as follows (see
eg. [1,2])  
\begin{eqnarray}
T^{iklm}& = & R^{iabl}{} {R^k_{~ab}}^m + {\star
R}^{iabl}{}{{\star R}^k_{~ab}}^m \nonumber \\
& = & R^{iabl}{} {R^k_{~ab}}^m + R^{iabm}{}
{R^k_{~ab}}^l - {1\over 2}g^{ik}R^{abcl}{}R_{abc}^{~~~m},
\end{eqnarray}
where $R_{iabl}$ is the Riemann tensor and $\star$ indicates the usual dual
operation 
\begin{equation}
{\star R}_{iabl} := {1\over 2}\eta_{iadc}{} R^{dc}_{~~bl}.
\end{equation}

The unusual properties (see eg. [2]) of the Bel--Robinson tensor
intrigued physicists and a large number of papers were devoted to the
understanding of its physical sense within the framework of the General
Relativity ({\bf GR}). Recently, in the Ref. [3] (see also [4]) the authors try
to connect this tensor with square of the energy--momentum and derive  from it
(by taking a suitable defined "square root") an energy--momentum tensor of the
gravitational field. In some special cases such square root really exists.
However, it seems that this is an incorrect physical idea. 

In order to understand the physical content of the Bel--Robinson tensor
correctly in {\bf GR} one should take into account the following,
fundamental facts: 
\begin{enumerate}
\item  The Bel--Robinson  tensor can be obtained  as a consequence of the
Bianchi identities independently of the Einstein equations.
\item  In the framework of the standard General Relativity ({\bf GR})
\footnote{The standard General Relativity has a very good experimental
confirmation; especially its main postulate --- Einstein Equivalence Principle
(see eg. [5,6,7]).}the gravitational field {\it has non--tensorial strenghts}
$\Gamma^i_{kl}$ and {\it  has no} (and cannot have any) energy--momentum tensor
but only {\it the so--called pseudotensors}.  It is a consequence of the {\it
Einstein Equivalence Principle} ({\bf EEP}). 
\item  The Bel--Robinson tensor {\it appears explicitly}  in the expansion of
the differences of the gravitational energy--momentum calculated in {\it normal
coordinates} {\bf NC(P)} by use of the canonical energy--momentum pseudotensor
$_E t_i^{~k}$. Namely, it is a part of the differences
\footnote{The analogous expansions were obtained by using other
energy--momentum pseudotensors of the gravitational field and also contain
Bel--Robinson tensor.  However, they are much more complicated; see e.g.
[8].} $_E t_i^{~k}(y) - _E t_i^{~k}(P)$. Here $y$ means normal
coordinates {\bf NC(P)} which have point {\bf P} as origin.
\end{enumerate}

In this paper, in Sec. II and in Sec. III,  we will consider the above three facts more intensively and
show that they {\it uniquely indicate} the link between Bel--Robinson tensor
and differences of the canonical gravitational energy--momentum calculated in
{\bf NC(P)}.\\
In Sec. IV we give conclusions and some remarks.
\section{The Bel--Robinson tensor and Bianchi identities}

The Bel--Robinson tensor is a special case of the so--called {\it Maxwellian
tensors}. The Maxwellian tensors generalize the symmetric
energy--momentum tensor of the electromagnetic field onto antisymmetric
tensor fields. The general method of construction of such  tensors  was 
developed in [9]. 
In the following we apply this  method to the Riemann tensor.

Let us consider Bianchi identities for the Riemann tensor $R_{iklm} = R_{lmik}
= - R_{kilm} = - R_{ikml}$ 
\begin{equation}
\nabla_{[a}{}R_{bc]de} \equiv
\nabla_a {}R_{bcde} + \nabla_b {}R_{cade} + \nabla_c {}R_{abde} \equiv
0
\end{equation}
and their non--vanishing contractions 
\begin{equation}
\nabla_a{} {R^{ab}_{~~cd}} \equiv 2\nabla_{[c}{}
R^b_{d]}.
\end{equation} 

We realize  that the identities (3)--(4) {\it possess Maxwellian structure} in
the indices $(a,b,c)$. 

Let us multiply (3) by $R^{bcd}_{~~~f}$. Then, after simple calculations we
get the new identities 
\begin{equation}
{R^{bcd}_{~~~f}} \nabla_b {}R_{acde} - {1\over
2}{R^{bcd}_{~~~f}} \nabla_a {} R_{bcde} \equiv 0.
\end{equation}
Then let us transpose the indices $f$ and $e$ in (5)
\begin{equation}
{R^{bcd}_{~~~e}}\nabla_b {} R_{acdf} - {1\over 2}
{R^{bcd}_{~~~e}}\nabla_a {}R_{bcdf} \equiv 0.
\end{equation}
The sum of (5) and (6) gives
\begin{eqnarray}
{R^{bcd}_{~~~f}} \nabla_b {}R_{acde} & + & {R^{bcd}_{~~~e}}\nabla_b
{}R_{acdf} \nonumber \\
&-& {1\over 2} \bigl({R^{bcd}_{~~~f}}\nabla_a {}R_{bcde} +
{R^{bcd}_{~~~e}}\nabla_a {}R_{bcdf}\bigr) \equiv 0.
\end{eqnarray}

Due to the identities (4) one can rewrite the identities (7) in the
following form 
\begin{eqnarray}
\nabla_b\bigl({R^{bcd}_{~~~f}}{}R_{acde} & + &
{R^{bcd}_{~~~e}}{}R_{acdf} - {1\over
2}\delta^b_a{R^{bcd}_{~~~e}}{}R_{bcdf}\bigr) \nonumber \\
& \equiv & 2{R_{ac}^{~~d}}_e
{}\nabla_{[d} R^c_{f]} + 2 {R_{ac}^{~~d}}_f {} \nabla_{[d} R^c_{e]}.
\end{eqnarray}

The {\it Bel--Robinson tensor} $T^b_{~aef}$ is easily indicated inside
parenthesis on the left hand side of the identities (8) which determine
the divergence of this tensor.

Thus, we see that the Bel--Robinson tensor and its divergence arise as a {\it consequence of the
Bianchi identities} (3) {\it and their contractions} (4) {\it only} and
they are  {\it neither connected with Einstein equations nor with the canonical formalism
of the energy--momentum in GR}.

However, by using of the Einstein equations 
\begin{equation}
R^i_k = \beta\bigl(T^i_k - {1\over
2}\delta^i_k T\bigr) =: \beta E^i_k,
\end{equation}
where $\beta = 8\pi G/ c^4$, one can rewrite the identities (8) in the
form 
\begin{equation}
\nabla_b {T^b_{~aef}} = 2\beta {R_{ac}^{~~d}}_e{}\nabla_{[d} E^c_{f]} + 2\beta
{R_{ac}^{~~d}}_f {}\nabla_{[d} E^c_{e]}.
\end{equation}

The equations (10) give {\it the link} between the divergence of the
Bel--Robinson tensor and {\bf GR}.

It follows from (10) that, in vacuum, 
\begin{equation}
\nabla_b {T^b_{~aef}} = 0.
\end{equation}

The dimensions of the components of the Bel--Robinson tensor are
$(length)^{(-)4}$; but it is a trivial fact that $1/\beta^2{}T^b_{~afe}$ have
dimensions of the energy--momentum square. This trivial fact was used in
Ref. 3 with the aim of  connecting  the Bel--Robinson tensor with square of an
energy--momentum tensor.
\section{The relation between  the Bel--Robinson tensor and the canonical 
energy--momentum in General Relativity}

The problem of the energy--momentum in General Relativity ({\bf GR}) was
intensively studied by many authors (see e.g. [10---17]).  The main results of
these investigations are the following: \footnote{We do not consider here the
so--called {\it quasilocal quantities} [18---22] because already for the Kerr
spacetime differently defined quasilocal quantities give different results
(see eg. [23]). Moreover, the term "quasilocal" is very obscure. We omit also {\it Lorentz hypothesis} (see eg. [24])
which is unsatisfactory from the physical point of view since it gives
for gravitational field an energy--momentum tensor which vanishes in
vacuum. But we think that Lorentz hypothesis is the best of the all
trials to attribute an energy--momentum tensor to gravitational field.}  
\begin{enumerate}
\item Owing to the non--tensorial character of the gravitational strengths
$\Gamma^i_{kl} = \bigl\{^i_{kl}\bigr\}$ the gravitational field in standard
{\bf GR} {\it has no} (and cannot have) any energy--momentum tensor. Any
attempt to introduce such a tensor leads us beyound standard {\bf GR}.
Moreover, it is speculative and contradicts {\bf EEP}.

From that it follows the {\it non--localizability} of the gravitational
energy--momentum. 
\item The best solution of the energy--momentum problem in standard {\bf GR}
seems to be given by the so--called {\it canonical energy--momentum
pseudotensor} $_E t_i^{~k}$ proposed for gravitational field by Einstein and
related to that pseudotensor, the {\it canonical, double index,
energy--momentum complex} 
\begin{equation}
_E K_i^{~k} := \sqrt{\vert g\vert}\bigl(T_i^k + _E 
t_i^{~k}\bigr),
\end{equation}
for matter and gravitation which satisfies
\begin{equation}
\sqrt{\vert g\vert}\bigl(T_i^k + _E t_i^{~k}\bigr) = _F
{U_i^{kl}}_{,l}.
\end{equation}
Here 
\begin{equation}
_F {U_i^{~kl}} = (-) _F {U_i^{~lk}} = \alpha{g_{ia}\over\sqrt{\vert
g\vert}}\bigl[(-g)\bigl(g^{ka} g^{lb} - g^{la}
g^{kb}\bigr)\bigr]_{,b}
\end{equation}
are {\it von Freud superpotentials}, $T_i^k$ are  the components of a
symmetric energy--momentum tensor of matter (the sources in the Einstein
equations) and $g$ is the determinant of the metric tensor; $,i$ or
$\partial_i$ denotes partial derivative. $\alpha = 1/2\beta = c^4/16\pi G$. 
\end{enumerate}

We have [25---27]
\begin{eqnarray}
_E t_i^{~k} & = & \alpha\biggl\{\delta_i^k
g^{ms}\bigl(\Gamma^l_{mr}{}\Gamma^r_{sl} -
\Gamma^r_{ms}{}\Gamma^l_{rl}\bigr)\nonumber \\
& + & g^{ms}_{~~,i}\bigl[\Gamma^k_{ms} -
{1\over 2}\bigl(\Gamma^k_{tp}{} g^{tp} - \Gamma^l_{tl}{}
g^{kt}\bigr)g_{ms}\nonumber \\
& - & {1\over 2}\bigl(\delta^k_s {}\Gamma^l_{ml} + \delta^k_m {}
\Gamma^l_{sl}\bigr)\bigr]\biggr\}.
\end{eqnarray}

The equations (13) can be obtained by rearranging of the Einstein equations
having $T^k_i$ as sources. 

From (13) there follow the {\it local or differential} conservation laws
\begin{equation}
\bigl[\sqrt{\vert g\vert}\bigl(T_i^k + _E t_i^{~k}\bigr)\bigr]_{,k} = 0,
\end{equation}
and, by using Stokes integral theorem, the {\it integral conservation laws}
\begin{equation}
\oint\limits_{\partial\Omega}\sqrt{\vert 
g\vert}\bigl(T_i^k + _E t_i^{~k}\bigr)d\sigma_k = 0.
\end{equation}

$\partial\Omega$ is the boundary of a four--dimensional, compact domain
$\Omega$, and $d\sigma_k$ is the  three--dimensional integration element
(see e.g. [17]). 

The components $_E t_i^{~k}$ of the Einstein canonical energy--momentum
pseudotensor of the gravitational field {\it neither form a tensor nor other
geometric object}.

Any attempt of the physical interpretation of the Bel--Robinson tensor in the
framework of {\bf GR} should take into account the connection of the
Bel--Robinson tensor and its divergence with Bianchi identities, the above two
fundamental facts and the next, more important fact as follows: {\it the
Bel--Robinson tensor appears explicitly in the expansion of the differences}
 
\begin{equation}
_E t_i^{~k}(y) - _E t_i^{~k}(P)
\end{equation}
{\it of the canonical energy--momentum} calculated in {\it normal
coordinates} [28---30] {\bf NC(P)}(And in analogic expansions
obtained when using other pseudotensors too  [8]).\footnote{Because $_E
t_i^{~k}(P) = {_E t_i^{~k}}_{,l}(P) =0,$ one can also speak about expansion of
$_E t_i^{~k}$ alone. But the differences $_E t_i^{~k}(y) - _E t_i^{~k}(P)$ have
deeper physical meaning; for example, they admit introduction of {\it
superenergy and supermomentum tensors} [31---37].} This fact gives 
{\it the most important connection} between the Bel--Robinson tensor
and {\bf GR} as follows.

In the {\bf NC(P)} $\{y^i\}$ having the point {\bf P} as their origin we have
[38] 
\begin{eqnarray}
_E t_i^{~k}(y) - _E t_i^{~k}(P) & = & _E
t_i^{~k}(y) = {1\over 2} {_E t_i^{~k}}_{,lm}(P) y^ly^m + R_3 \nonumber \\
& = & {\alpha\over 9}\bigl[T^k_{~ilm}(P) + P^k_{~ilm}(P) - {1\over 2}\delta_i^k
R^{abc}_{~~~l} (P) \bigl(R_{abcm}(P) + R_{acbm}(P)\bigr) \nonumber \\
& + & 2\delta_i^k R_{(l\vert g}(P){} R^g_{~\vert m)}(P) - 3 R_{i(l\vert}(P){}
R^k_{~\vert m)}(P) + R^k_{~gi(l\vert}(P){} R^g_{~\vert m)}(P)\nonumber \\
& + & R^k_{~ig(l\vert}(P){}R^g_{~\vert m)}(P)\bigr] y^ly^m + R_3.
\end{eqnarray} 

In the formula (19) $R_3$ is the remainder of the third order, while 
\begin{equation}
T^k_{~ilm}:= R^{kab}_{~~~l}{} R_{iabm} + R^{kab}_{~~~m} 
{} R_{iabl} - {1\over 2}\delta^k_i R^{abc}_{~~~l}{} R_{abcm}
\end{equation}
are the {\it Bel--Robinson} tensor components, and 
\begin{equation}
P^k_{~ilm} := R^{kab}_{~~~l}{}R_{ibam} + R^{kab}_{~~~m}{} R_{ibal} - {1\over
2}\delta^k_i R^{abc}_{~~~l}{} R_{acbm}
\end{equation}
are components of the tensor which is closely related to the Bel--Robinson
tensor.

By using the Einstein equations in the form (9), one can rewrite (19) to the
form 
\begin{eqnarray}
_E t_i^{~k}(y) - _E t_i^{~k}(P) & = & _E t_i^{~k}(y) =
{\alpha\over 9}\bigl[T^k_{~ilm} (P) + P^k_{~ilm} (P) \nonumber \\
& - & {1\over 2}\delta_i^k R^{abc}_{~~~l}(P)\bigl(R_{abcm}(P) +
R_{acbm}(P)\bigr)\nonumber \\
& + & 2\delta_i^k \beta^2 E_{(l\vert g}(P){} E^g_{~\vert m)}(P) - 3\beta^2
E_{i(l\vert}(P){}E^k_{~\vert m)}(P) + \beta R^k_{~gi(l\vert}(P){}E^g_{~\vert
m)}(P)\nonumber \\
& + & \beta R^k_{~ig(l\vert}(P) {}E^g_{~\vert m)}(P)\bigr] y^ly^m + R_3.
\end{eqnarray}

In vacuum we have from (22) 
\begin{eqnarray}
_E t_i^{~k}(y) & - & _E t_i^{~k}(P)  = _E t_i^{~k}(y) = {\alpha\over
9}\bigl[T^k_{~ilm}(P) + P^k_{~ilm}(P) \nonumber \\
& - & {1\over 2} \delta_i^k R^{abc}_{~~~l}(P)\bigl(R_{abcm}(P)
+ R_{acbm}(P)\bigr)\bigr]y^ly^m + R_3 \nonumber \\
& = & {4\alpha\over 9}\bigl[R^{k(ab)}_{~~~~~(l\vert}(P) {} R_{iab\vert m)}(P) -
{1\over 2}\delta_i^k{} R^{a(bc)}_{~~~~~l}(P) {} R_{abcm}(P)\bigr]y^ly^m + R_3.
\end{eqnarray}
We see from the above formulas (19)--(23) that the Bel--Robinson tensor
{\it really appears} in the expansion of the differences (18).

Years ago, by using the expansion (23), we have shoved [38] that the
infinitesimal differences $\triangle P_a$ of the free gravitational
energy--momentum calculated in {\bf NC(P)} are proportional to the components 
$T^0_{~a00}$ of the Bel--Robinson tensor multiplied by $\alpha = c^4/16\pi G$.
\section{Conclusion}
The Bel--Robinson tensor follows from  the Bianchi identities {\it as the
Maxwellian tensor} for the Riemann tensor $R_{iklm}$ and, within the framework of the standard  {\bf GR}, {\it it
can be connected with the differences of the canonical gravitational
energy--momentum calculated in} {\bf NC(P)}. It is easily seen from Sec.
II, from the formulas (19)--(23) and from the Ref. 38. These facts give
the most natural (and correct) physical interpretation of this tensor in
the framework of the {\bf GR}.
 
Although the gravitational field $\Gamma^i_{kl}$ has no energy--momentum
tensor in standard {\bf GR}, one can easily introduce there the so--called {\it canonical superenergy
tensor}  [31---37] for this field. 
The method of construction of the canonical superenergy tensor for
gravitational field uses the expansion (19) and some kind of averaging. The Bel--Robinson tensor
multiplied by $\alpha = c^4/16\pi G$ is the main, "Maxwellian part"  of such a
tensor. 

The canonical superenergy tensor for gravitational field does not vanish
in vacuum; so, it gives us, for example, a very useful tool for local analysis of
the gravitational radiation [34].

The idea of superenergy  and its tensor is {\it universal } and easy to
generalize onto matter field too; for example, one can easily introduce {\it
the canonical superenergy tensor for  matter} [31---35]. 

On the "superenergy level" one can easily introduce {\it the canonical
angular supermomentum tensors} for matter and gravitation [36,37] as well.

The {\it canonical superenergy tensors} of gravitation and matter and the
{\it canonical angular supermomentum tensors} of gravitation and matter have
much better geometrical and physical properties than the canonical objects
from which they were obtained. Also, the integral superenergetic quantities
have better properties than corresponding integral energetic quantities;
especially, the superenergetic integrals  have better convergence in
asymptotically flat spacetimes (at spatial or null infinity).

Finally, the canonical superenergy and angular supermomentum tensors
give a very good tool for local (and also to global) analysis of the   
gravitational and matter fields within the framework of the standard 
{\bf GR}.\hfill\break
\bigskip
\centerline{\bf Acknowledgments}
\medskip
I would like to thank Dr M.P. D\c{a}browski for his help in preparation
English version of this paper.

\end{document}